\documentclass[aps,epsf,eqsecnum,feynmp,graphics,amsfonts, amssymb, latexsym, amsmath]{revtex4}
\usepackage[latin1]{inputenc}
\usepackage{graphics}
\usepackage{showlabels}
\usepackage{amsfonts, amssymb, latexsym, amsmath}
\def\aut#1{#1}
\def\comment#1{}
\begin{document}
\title{Variational Perturbation Theory for
Summing Divergent Non-Borel-Summable Tunneling Amplitudes}
\author{B.~Hamprecht and H.~Kleinert}
\address{Institut für Theoretische Physik, Freie Universität Berlin,\\
Arnimallee 14, D-14195 Berlin, Germany\\
{\scriptsize e-mails: bodo.hamprecht@physik.fu-berlin.de}
{\scriptsize e-mails: hagen.kleinert@physik.fu-berlin.de}}
\begin{abstract}
We present a method for
evaluating divergent
non-Borel-summable
 series
by
 an analytic continuation
 of variational perturbation
 theory. We demonstrate the power of the method
by an application to the exactly known partition function
of the anharmonic oscillator
in zero space-time dimensions.
In one space-time dimension
we
derive
the imaginary part of the
 ground state energy of
 the anharmonic oscillator for {\em all\/} negative
 values of
 the coupling constant $g$, including the
non-analytic tunneling regime at
small $-g$.
As a
highlight
 of the theory
we retrieve
 the divergent
perturbation expansion from
the  action
of the critical bubble
and the contribution
of the higher  loop
fluctuations around the bubble.
\end{abstract}
\maketitle
\section{Introduction}
None of
 the presently
known resummation schemes \cite{Resumm,Verena}
is able to deal with non-Borel-summable series.
Such series arise
in
 the
 theoretical description
of many
important physical phenomena,
in particular tunneling
processes.
In the path integral, these are dominated by
non-perturbative contributions coming from
nontrivial classical solutions
called {\em critical bubbles\/} \cite{Langer,PI}
or {\em bounces\/} \cite{Coleman},
and fluctuations around these.

Any Borel-summable
series becomes non-Borel-summable
if
 the expansion parameter, usually some coupling constant $g$,
is continued to negative values.
Here we show that
non-Borel-summable
series  can be evaluated with any desired accuracy
 by an
analytic continuation
of
 variational perturbation
 theory
 \cite{PI,Verena} in
 the complex $g$-plane.
This implies that
 variational perturbation
 theory
can give us information on
non-perturbative properties of the theory.

Variational perturbation
 theory
has a long history
\cite{refs,finiteg,KleinertJanke,Guida}. It is based on the introduction of a
dummy variational
 parameter $ \Omega $
on which
the full perturbation expansion does not depend,
while the truncated expansion
does. An optimal $ \Omega $ is selected
by the principle of minimal sensitivity \cite{Stevenson},
 requiring the
quantity of interest to be stationary as function of the
variational parameter. The optimal
$ \Omega $
 is usually
taken from a zero of
 the derivative with
respect to $ \Omega $.
If the first derivative has no zero,
a zero of the second derivative is chosen.
For Borel-summable series, these zeros are always real, in contrast
to statements
in
the literature
 \cite{Neveu,RULES,Braaten,Ramos}
which have proposed
the use of complex
zeros. Complex zeros produce
in general  wrong results
for Borel-summable series, as was recently shown
in Ref.~\cite{AntiBraaten}.

The purpose of this paper is to show that there
does exist
 a wide range of applications
of complex zeros if one
wants to resum
 non-Borel-summable series, which have so far remained intractable.
These arise typically
in tunneling problems,
and we
shall
see
 that variational perturbation
theory provides us with an
efficient method for evaluating
these series,
 rendering
their real
and imaginary parts
with any desirable accuracy,
if only enough perturbation coefficients are available.
An important problem which had to be solved
is the specification of the proper
choice
the optimal zero from  the many possible candidates
existing in higher orders.
A non-Borel-summable series
is associated with a function which has
 an essential singularity at the origin
in the complex $g$-plane,
which is the starting point of a
 left-hand cut. Near the
 tip of the cut,
the imaginary part of the function  approaches zero
rapidly like $\exp({-\alpha/|g|})$
for $g \to 0-$.
If the variational approximation
is plotted against $g$ with
an enlargement factor $\exp({\alpha/|g|})$,
oscillations
become visible near $g=0$.
The choice of the optimal
 complex zeros
of the derivative with respect to the variational parameter
is fixed
 by the
requirement
of obtaining, in each order,
the least oscillating
imaginary part when approaching
 the tip of
 the cut. We may call this selection rule the
{\em principle of minimal
sensitivity and oscillations\/}.

In Section \ref{@sec2},
we shall explain and test
 the new principle
 on the exactly known partition function
$Z(g)$ of the anharmonic oscillator
in zero space time dimensions.
%
In Section \ref{@sec3}, we apply the method to the
critical-bubble regime of small $-g$ of the anharmonic oscillator
and find the action of the critical bubble and the
corrections caused by the  fluctuations around it.
In Section \ref{@sec4} we present
yet another
method of calculating the
properties of the
critical-bubble regime.
This method
is
restricted to quantum mechanical systems.
Its results for the anharmonic oscillator
 give
 more evidence
for the correctness of the general method
of Sections \ref{@sec2} and \ref{@sec3}.
\section{Test of
 Resummation of
Non-Borel-summable Expansions}
\label{@sec2}
The partition function
$Z(g)$ of the anharmonic oscillator
in zero space-time dimensions is
\begin{align}
\label{FOKKER}
Z(g) = \frac{1}{ \sqrt{\pi}}\; \int_{-\infty}^\infty
 \exp{(-x^2/2-g\ x^4/4)}\ dx=\frac{\exp{(1/8g)} }
{\sqrt{4\pi g}}K_{1/4}(1/8g)\ ,
\end{align}
where $K_{ \nu }(z)$ is the modified Bessel function.
For small $g$, the function $Z(g)$ has a divergent
Taylor
series expansion, to be called
{\em weak-coupling expansion\/}:
\begin{align}
\label{FP-WEAK}
Z_{\rm weak}^{(L)}(g) = \sum_{l=0}^L\; a_l\;g^l, \qquad \!\!\!\!\!\mbox{with }\ a_l=(-1)^l\ \frac{\Gamma(2l+1/2)}{l!\sqrt \pi} .
\end{align}
For $g<0$, this is
non-Borel-summable.
For large  $|g|$ there exists a convergent {\em strong-coupling expansion\/}:
\begin{align}
\label{FP-STRONG}
Z_{\rm strong}^{(L)}(g) =g^{-l/4}\ \sum_{l=0}^L\; b_l\;g^{-l/2}, \qquad\!\!\!\!\!
 \mbox{with }\ b_l=(-1)^l\ \frac{\Gamma(l/2+1/4)}{2 l!\sqrt \pi}.
\end{align}
%
As is obvious from the integral representation (\ref{FOKKER}),
$Z(g)$
obeys
the second-order differential equation
\begin{align}
\label{DGL}
16 g^2 Z''(g)+4(1+8g)Z'(g)+3 Z(g)=0,
\end{align}
which has two independent solutions.
One of them is
 $Z(g)$, which  is finite for $g>0$ with $Z(0)=a_0$.
 The
weak-coupling
coefficients $a_l$ in
(\ref{FP-WEAK})
can be obtained
by inserting
into
 (\ref{DGL})
 the Taylor series
and comparing coefficients.
The result is the
recursion relation
\begin{align}
\label{REC}
a_{l+1}=-\frac{16l(l+1)+3}{4(l+1)}\, a_l.
\end{align}

A similar recursion relation
can be derived for the strong-coupling coefficients $b_l$ in Eq.~(\ref{FP-STRONG}).
We observe that
 the two independent solutions
$Z(g)$
of (\ref{DGL})
behave like $Z(g) \propto g^\alpha$ for $g \to \infty$ with the powers
$\alpha=-1/4$ and $-3/4$. The
function
(\ref{FOKKER})
has
 $\alpha=-1/4$.
It is convenient to remove the leading power
from $Z(g)$ and define a function $\zeta(x)$
such that  $Z(g) = g^{-1/4}\ \zeta(g^{-1/2})$.
The Taylor coefficients of $\zeta(x)$
 are the strong-coupling coefficients $b_l$ in Eq.~(\ref{FP-STRONG}).
The function $ \zeta (x)$  satisfies
 the differential equation
and initial conditions:
\begin{align}
\label{DGL2}
4\zeta''(x)-2x\zeta'(x)-\zeta(x) &=0,~~~~\mbox{with}~~
\zeta(0)=b_0 ~~\mbox{and}~~ \zeta'(0)=b_1 .
\end{align}
The
Taylor coefficients $b_l$  of $\zeta(x)$
satisfy the recursion relation
\begin{align}
\label{REC2}
b_{l+2}=\frac{2l+1}{4(l+1)(l+2)}b_{l} \,.
\end{align}
Analytic continuation of $Z(g)$ around $g=\infty$ to
 the left-hand cut gives:
\begin{align}
\label{CUT}
Z(-g)&=(-g)^{-1/4} \zeta((-g)^{-1/2})\\
&=(-g)^{-1/4} \sum_{l=0}^\infty b_l(-g)^{-l/2} \exp{\left[-\frac{i\pi}{4}(2l+1)\right] } \qquad \mbox{for $g>0$},
\end{align}
so that we find an imaginary part
\begin{align}
\label{CUTi}
{\rm Im} \, Z(-g)&= -(4g)^{-1/4}  \sum_{l=0}^\infty b_l(-g)^{-l/2} \sin{\left[
-\frac{i\pi}{4}(2l+1)\right] }\\
&=-(4g)^{-1/4}\   \sum_{l=0}^\infty \beta_l(-g)^{-l/2}  \ ,
\end{align}
where
\begin{align}
\label{CUT2}
\beta_0&=b_0,~~~~ \beta_1=b_1,~~~~ \beta_{l+2}=-\frac{2l+1}{4(l+1)(l+2)}\beta_{l}\ .
\end{align}
It is easy to show that
\begin{equation}
\sum_{l=0}^\infty \beta_l x^l=\zeta(x)\exp{(-x^2/4)},
 \label{@}\end{equation}
so that
\begin{align}
\label{CUT-STRONG}
{\rm Im}\, Z(-g)=-\frac{1}{\sqrt 2}\, g^{-1/4}\  \exp{(-1/4g)}\
 \sum_{l=0}^\infty \; b_l\;g^{-l/2}\ .
\end{align}
From this we may re-obtain
the weak-coupling coefficients $a_l$ by means of the dispersion relation
\begin{align}
\label{DISP}
 Z(g)=& -\frac{1}{\pi}\int_0^\infty \frac{{\rm Im}\, Z(-z)}{z+g}\,dz\\
=& \frac{1}{\pi \sqrt 2}\ \sum_{j=0}^\infty \ b_j \int_0^\infty
\frac{  \exp{(-1/4z)}\  z^{-j/2-1/4 }}{z+g}\,dz.
\end{align}
Indeed, replacing $1/(z+g)$ by $\int_0^\infty \exp{(-x(z+g))}\,dx$,
and expanding  $\exp{(-x\,g)}$ into a power series, all integrals can be evaluated to yield:
\begin{align}
\label{DISP-2}
 Z(g)=& \frac{1}{\pi}\ \sum_{j=0}^\infty \ 2^j b_j \ \sum_{l=0}^\infty \ (-g)^l \Gamma(l+j/2+1/4)\ .
\end{align}
Thus we find
for  the weak-coupling coefficients $a_l$ an expansion in terms of
the strong-coupling coefficients
\begin{align}
\label{DISP-3}
 a_l=& \frac{(-1)^l}{\pi}\ \sum_{j=0}^\infty \ 2^j b_j \ \Gamma(l+j/2+1/4).
\end{align}
Inserting $b_j$ from
Eq.~(\ref{FP-STRONG}), this becomes
\begin{align}
\label{DISP-3}
 a_l=
 \frac{(-1)^l}{2\pi^{3/2}}\ \sum_{j=0}^\infty \  \frac{2^j (-1)^j}{j!}\ \Gamma(j/2+1/4)\Gamma(l+j/2+1/4)
=(-1)^l\ \frac{\Gamma(2l+1/2)}{l!\sqrt \pi}\ ,
\end{align}
coinciding with
(\ref{FP-WEAK}).

Variational perturbation theory is a well-established
 method
for obtaining  convergent
strong-coupling expansions from divergent
weak-coupling expansions  in
quantum-mechanical
systems such as
 the anharmonic oscillator
\cite{JK,PI}
as well as in quantum field theory \cite{strong,Verena}.
We have seen in Eq.~(\ref{CUT}), that
the
 strong-coupling expansion
can easily be continued analytically
to negative
$g$.  This continuation can, however, be used for an evaluation only for sufficiently
large $|g|$
where
the
 strong-coupling expansion
converges.
In the
 tunneling regime
near the tip of the left-hand cut,
the expansion diverges.
In this paper
we shall find that
an evaluation of
the weak-coupling expansion
according to the rules of variational perturbation theory
continued
into the complex plane
 gives extremely good results on the entire
left-hand cut with a fast convergence even near the tip
at $g=0$.

\begin{figure}[htb]
\begin{center}
\setlength{\unitlength}{1cm}
\begin{picture}(19,12.5)
\put(0.5,0.5){\scalebox{.8}[.8]{\includegraphics*{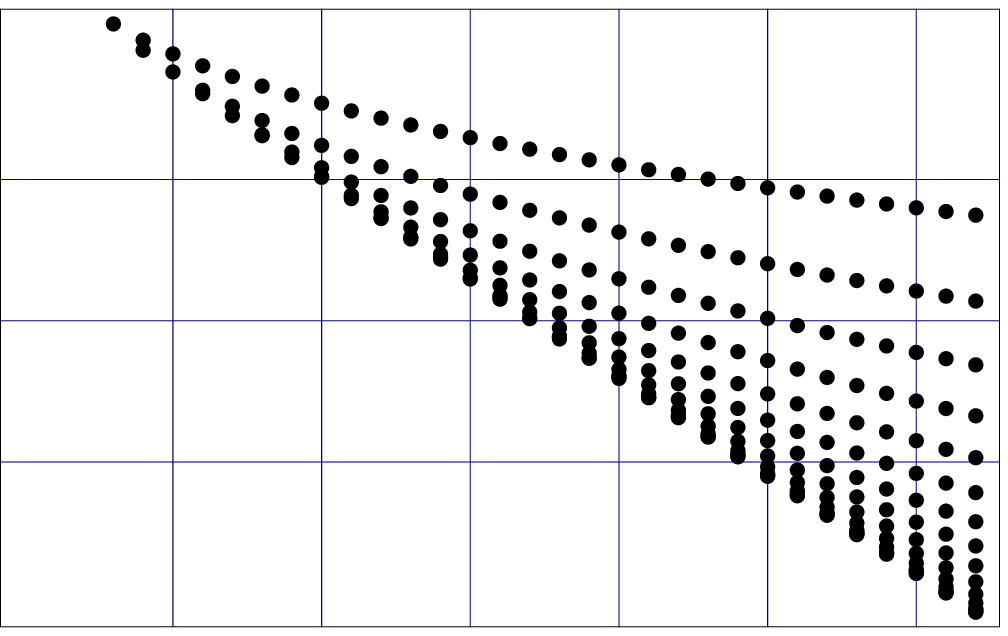}}}
\put(0.5,7){\scalebox{.8}[.8]{\includegraphics*{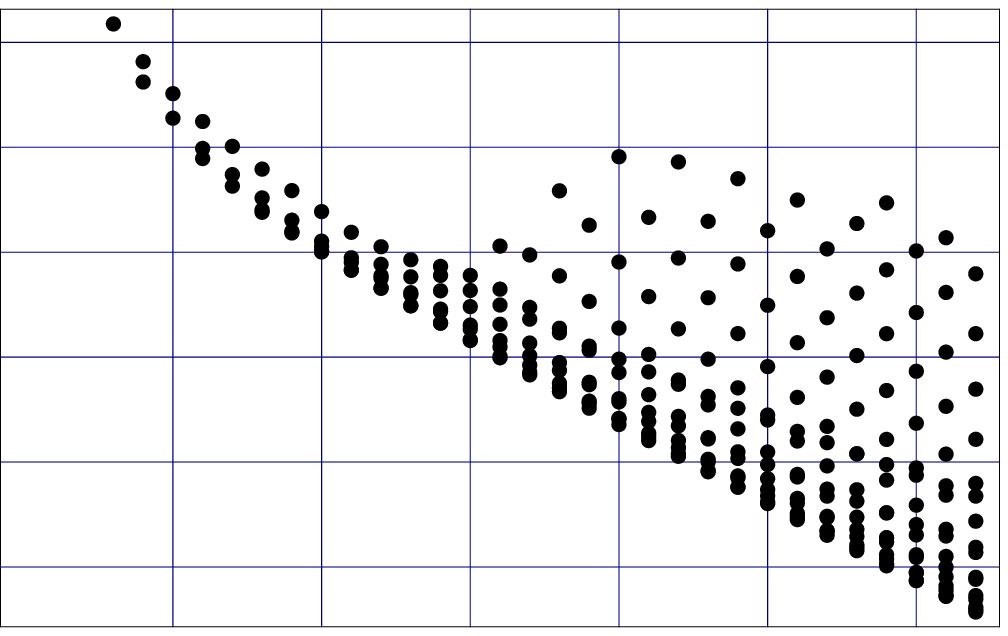}}}
\put(10,0.5){\scalebox{.8}[.8]{\includegraphics*{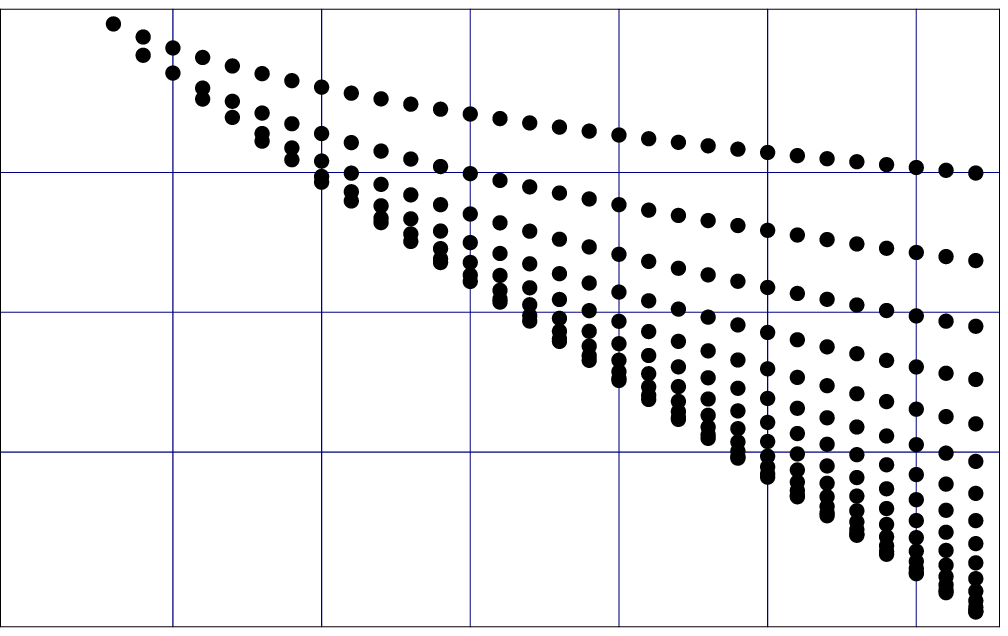}}}
\put(10,7){\scalebox{.8}[.8]{\includegraphics*{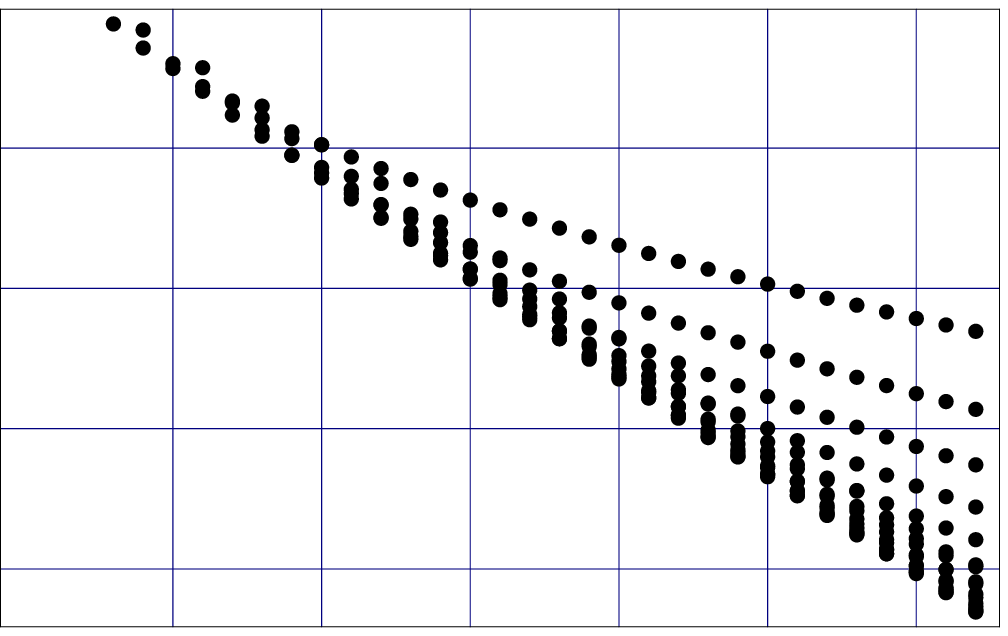}}}
\put(2.,7.7){$g=-.01$}
\put(11.5,7.7){$g=-.1$}
\put(2.,1.4){$g=-1$}
\put(11.41,1.4){$g=-10$}
\put(2.95,6.7){$10$}
\put(5.35,6.7){$20$}
\put(7.75,6.7){$30$}
\put(8.55,6.7){$L$}
\put(2.95,.2){$10$}
\put(5.35,.2){$20$}
\put(7.75,.2){$30$}
\put(8.55,.2){$L$}
\put(12.45,6.7){$10$}
\put(14.85,6.7){$20$}
\put(17.25,6.7){$30$}
\put(18.05,6.7){$L$}
\put(12.45,.2){$10$}
\put(14.85,.2){$20$}
\put(17.25,.2){$30$}
\put(18.05,.2){$L$}
\put(-.1,10.85){$-20$}
\put(-.1,9.1){$-30$}
\put(-.1,7.4){$-40$}
\put(9.4,10.87){$-10$}
\put(9.4,8.54){$-20$}
\put(-.1,4.17){$-10$}
\put(-.1,1.76){$-20$}
\put(9.4,4.15){$-10$}
\put(9.4,1.88){$-20$}
\end{picture}
\caption[NB-I]{Logarithm of deviation of the variational results
from  exact values $\log{|Z_{\rm var}^{(L)}-Z_{\rm exact}|}$
 plotted against the  order $L$ for different $g<0$ in the
non-Borel-summable region. All complex optimal $ \Omega $'s have been used.}
\label{NB-I}
\end{center}
\end{figure}

 The $L$th variational approximation to $Z(g)$ is given by
(see \cite{strong,Verena})
\begin{equation}
\label{FP-VAR}
Z_{\rm var}^{(L)}(g,\Omega)
 = \Omega^{p} \ \sum_{j=0}^L \left(\frac{g}{\Omega^q}\right)^j \epsilon_j(\sigma),~~~
\end{equation}
with
\begin{equation}
\label{FP-s}
\sigma\equiv \Omega^{q-2}(\Omega^2-1)/g\,,
\end{equation}

where $q=2/\omega=4$, $p =-1$ and
\begin{align}
\label{FP-EPS}
\epsilon_j(\sigma) = \sum_{l=0}^j a_l \binom{(p-lq)/2}{j-l} (-\sigma)^{j-l}\,.
\end{align}
To apply the principle of minimal sensitivity, the
zeros of the derivative of $Z_{\rm var}^{(L)}(g,\Omega)$ with respect to $\Omega$ are needed. They are given by the
zeros of the polynomials in $ \sigma $:
\begin{align}
\label{FP-DERIV}
P^{(L)}(\sigma) = \sum_{l=0}^L a_l (p-lq+2l-2L) \binom{(p-lq)/2}{L-l} (-\sigma)^{L-l}
=0,
\end{align}
since it can be shown
 \cite{JKsig} that the derivative depends only on $ \sigma $:
\begin{align}
\label{FP-DERIV-II}
\frac{d Z_{\rm var}^{(L)}(g,\Omega)}{d\Omega} =\Omega^{p-1} \left(\frac{g}{\Omega^q}\right)^L P^{(L)}(\sigma)\,.
\end{align}
\begin{figure}[htp!]
\begin{center}
\setlength{\unitlength}{1cm}
\begin{picture}(19,6.5)
\put(0.5,0.5){\scalebox{.8}[.8]{\includegraphics*{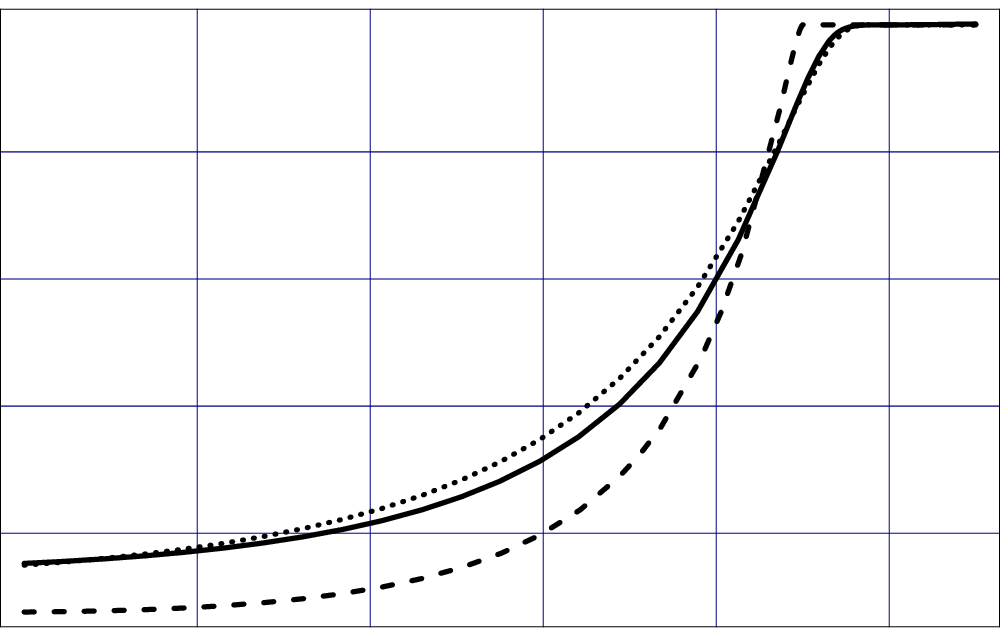}}}
\put(10,0.5){\scalebox{.8}[.8]{\includegraphics*{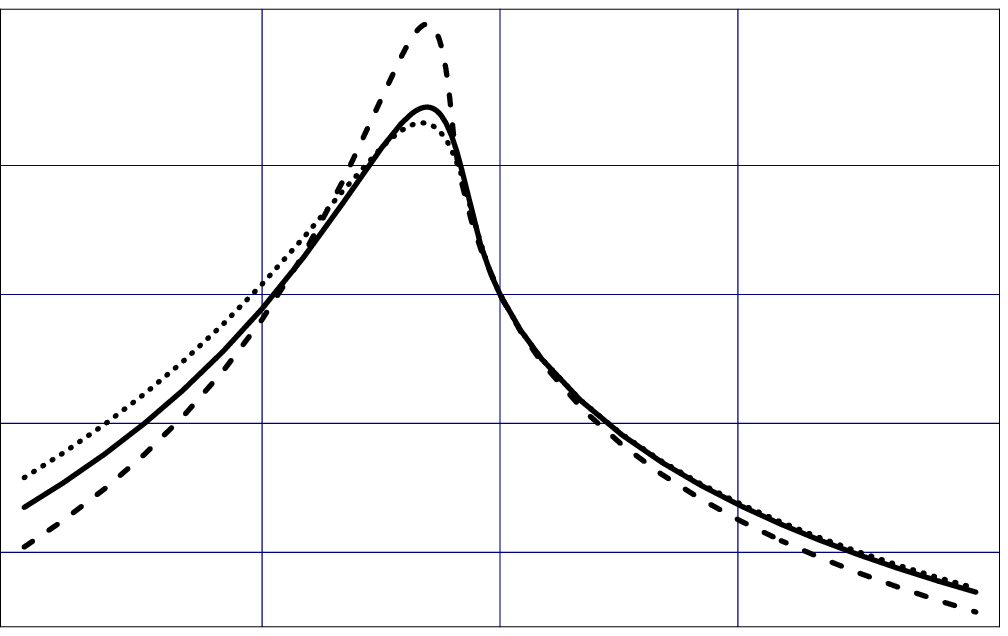}}}
\put(8.5,0.2){$g$}
\put(1.73,0.2){$-.8$}
\put(3.73,3.8){$Z(g)$}
\put(14.73,3.8){$Z(g)$}
\put(4.6,0.2){$-.4$}
\put(7.64,0.2){$0$}
\put(18,0.2){$g$}
\put(11.8,0.2){$-.5$}
\put(14,0.2){$0$}
\put(15.85,0.2){$.5$}
\put(-.1,3.25){$-.2$}
\put(-.1,1.2){$-.4$}
\put(9.8,3.15){$1$}
\put(9.7,1){$.8$}
\end{picture}
\caption[O2]{
Result of the 1st-
and 2nd-order
calculation for the
non-Borel-summable region of $g<0$, where the function has a
cut with non-vanishing imaginary part:
imaginary (left) and
 real parts (right) of $Z_{\rm var}^{(1)}(g)$ (dashed curve) and
$Z_{\rm var}^{(2)}(g)$ (solid curve) are plotted against $g$
and compared
with the
 exact values of the partition function (dotted curve).
The root of (\ref{FP-s}) giving the optimal variational parameter $ \Omega $ has been chosen to reproduce
the weak-coupling result near $g=0$.
}
\label{O2}
\end{center}
\end{figure}
Consider in more detail  the lowest non-trivial order with $L=1$.
 From Eq.~(\ref{FP-DERIV}) we obtain
\begin{align}
\label{Zero}
\sigma= &\frac{5}2,~~~~~~ {\rm corresponding~to}~~~~ \Omega= \frac{1}{2}\Big(1 \pm \sqrt{1 + 10g}\Big)\ .
\end{align}
In order to ensure that  our method  reproduces
 the weak-coupling result for small $g$, we have to take the positive sign in front of the square root. In
Fig.~\ref{O2} we have plotted $Z_{\rm var}^{(1)}(g)$ (dashed curve)
and $Z_{\rm var}^{(2)}(g)$ (solid curve) and compared these
with the exact result (doted curve) in the tunneling regime.
The agreement is quite good
even at these low orders \cite{Tunn}.
Next we study the behavior of $Z_{\rm var}^{(L)}(g)$ to higher orders $L$. For
selected coupling values
in the
non-Borel-summable region, $g=-.01,\ -.1,\ -1,\ -10$,
 we want to see the error as a function of the order.
We want to find  from this model system
the rule for selecting systematically the
best zero of $P^{(L)}(\sigma)$ solving Eq.~(\ref{FP-DERIV}),
which leads to
the optimal value
of the variational parameter $\Omega$.
For this purpose  we plot the variational results of all zeros. This
is shown in Fig.~\ref{NB-I}, where the logarithm of the deviations
from the exact value is plotted against the order $L$.
The outcome of different zeros cluster  strongly near the best value.
Therefore, choosing any zero out of the middle of the cluster is reasonable, in particular, because it does not depend on the knowledge of the exact solution,
so that this rule may be taken over to realistic cases.
\begin{figure}[htp!]
\begin{center}
\setlength{\unitlength}{1cm}
\begin{picture}(15,9)
\put(0,.7){\scalebox{1.25}[1.25]{\includegraphics*{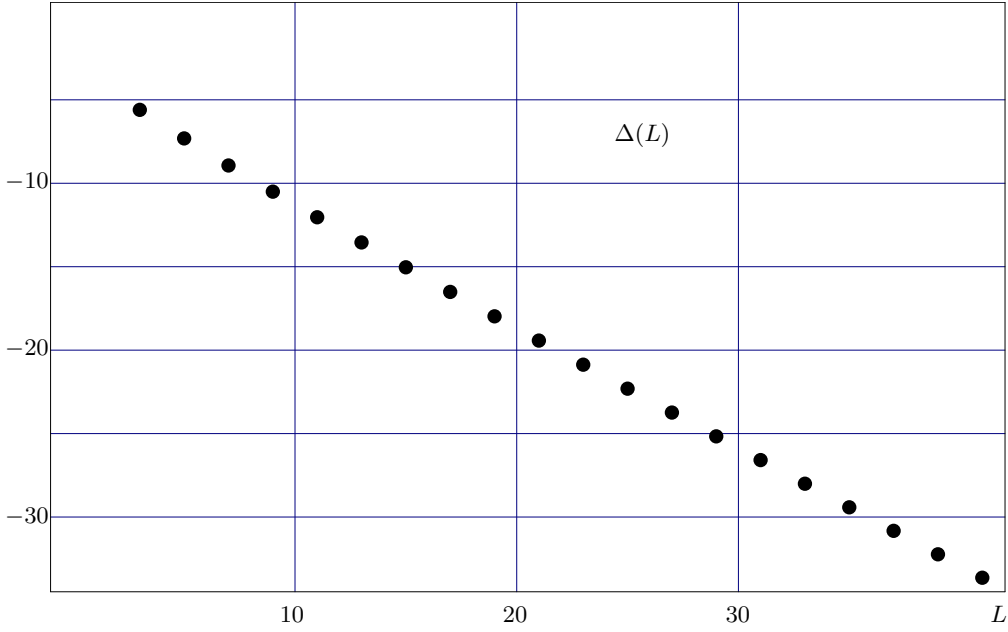}}}
\put(7.5,6.7){$ \Delta (L)$}
\put(12.5,0.3){$L$}
\put(3.05,0.3){$10$}
\put(6.02,0.3){$20$}
\put(8.98,0.3){$30$}
\put(-.6,6.08){$-10$}
\put(-.6,3.85){$-20$}
\put(-.6,1.62){$-30$}
\end{picture}
\caption[NB-II]{Logarithm of deviation of
 variational results from
exactly known value $\Delta(L)=\log{|Z_{\rm var}^{(L)}-Z_{exact}|}$,
 plotted against the  order $L$ for  $g=10$ in Borel-summable region.
The real positive
 optimal $ \Omega $
 have been used.
There is only one real zero of the first derivative
in every odd order $L$
and none for even orders.  There is
excellent convergence $\Delta(L)\simeq 0.02\exp{(-0.73L)}$
for $L\to\infty$.
}
\label{NB-II}
\end{center}
\end{figure}
\begin{figure}[htp!]
\begin{center}
\setlength{\unitlength}{1cm}
\begin{picture}(19,6.5)
\put(0.5,0.5){\scalebox{.8}[.8]{\includegraphics*{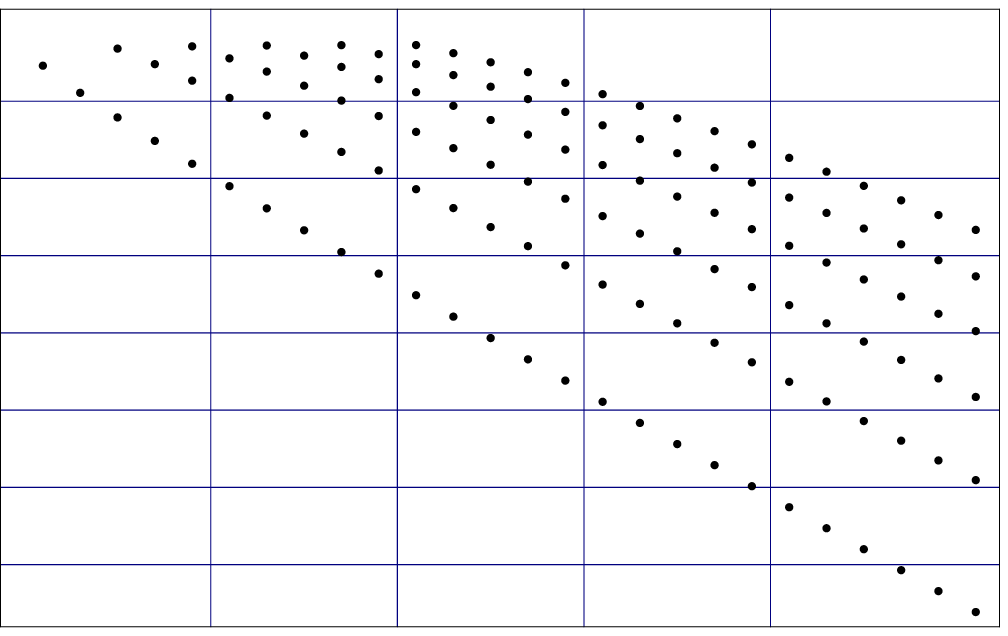}}}
\put(10,0.5){\scalebox{.8}[.8]{\includegraphics*{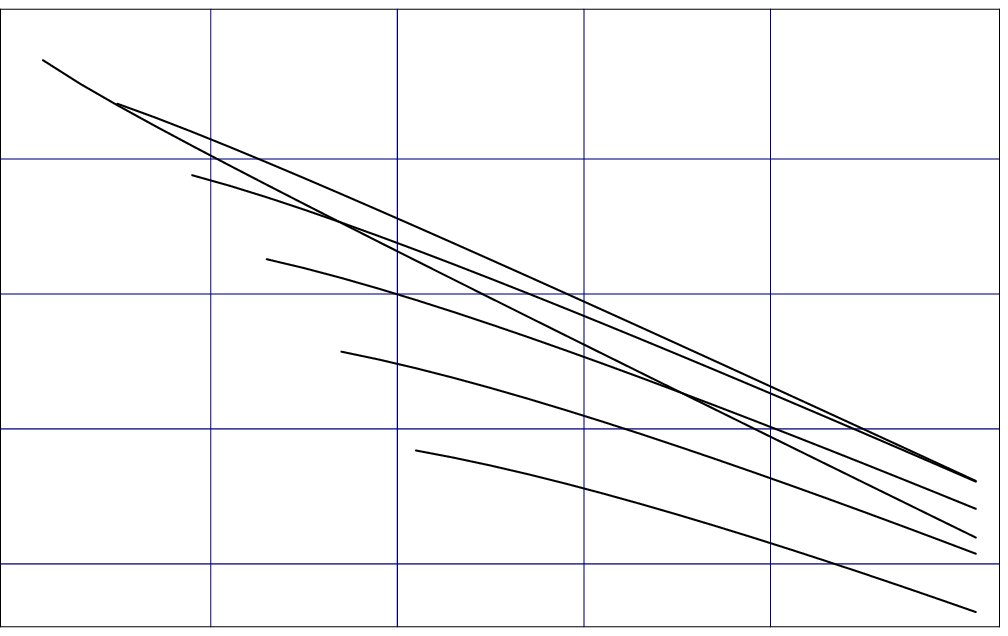}}}
\put(8.5,0.2){$L$}
\put(2.03,0.2){$10$}
\put(3.55,0.2){$20$}
\put(5.07,0.2){$30$}
\put(6.59,0.2){$40$}
\put(18,0.2){$L$}
\put(11.53,0.2){$10$}
\put(13.05,0.2){$20$}
\put(14.57,0.2){$30$}
\put(16.09,0.2){$40$}
\put(-.09,4.08){$-10$}
\put(-.09,2.82){$-20$}
\put(-.09,1.58){$-30$}
\put(9.4,4.28){$-10$}
\put(9.4,3.18){$-20$}
\put(9.4,2.08){$-30$}
\put(9.4,.98){$-40$}
\put(2.6,1.2){$\Delta_r$}
\put(12.1,1.2){$\Delta_a$}
\put(5.8,2){$0$}
\put(6.41,2.68){$4$}
\put(7,3){$8$}
\put(7.21,3.3){$12$}
\put(7.5,3.65){$16$}
\put(7.86,3.9){$20$}
\put(10.15,5.08){$0$}
\put(11.4,4.63){$4$}
\put(11.37,4.1){$8$}
\put(11.8,3.44){$12$}
\put(12.4,2.7){$16$}
\put(12.89,1.9){$20$}
\end{picture}
\caption[COEFF]{Relative logarithmic error $\Delta_r=\log{|1-b_l^{(L)}/b_l^{\rm (exact)}|}$ on the left,
and the absolute logarithmic error
$\Delta_a=\log{|b_l^{(L)}-b_l^{\rm (exact)}|}$ on the right,
plotted for some strong-coupling coefficients
$b_l$ with $l=0,4,8,12,16,20$
 against the order $L$.}
\label{COEFF}
\end{center}
\end{figure}
We wish to emphasize, that for the Borel-summable
domain with $g>0$,
 variational perturbation theory has the usual
fast convergence in this model.
In fact, for $g=10$, probing deeply into the
strong-coupling domain, we find rapid convergence like $\Delta(L)\simeq 0.02\exp{(-0.73L)}$ for $L\to\infty$, where $\Delta(L)=\log{|Z_{\rm var}^{(L)}-Z_{\rm exact}|}$ is the logarithmic error as a function of the order $L$. This
is shown in Fig.~\ref{NB-II}.
Furthermore, the strong-coupling coefficients $b_l$
of Eq.~(\ref{FP-STRONG}) are reproduced quite satisfactorily.
Having solved $P^{(L)}(\sigma)=0$ for $\sigma$,
we obtain $\Omega^{(L)}(g)$ by solving Eq.~(\ref{FP-s}). Inserting this
and (\ref{FP-EPS}) into (\ref{FP-VAR}), we bring $g^{1/4}\ Z_{\rm var}^{(L)}(g)$ into a form suitable for expansion in powers of $g^{-1/2}$. The expansion coefficients are the strong-coupling coefficients $b_l^{(L)}$ to order $L$. In Fig.~\ref{COEFF} we have plotted the logarithms of their absolute
and relative errors over the order $L$,
and find very good convergence, showing that
variational perturbation theory
works well for
our test-model $Z(g)$.

\begin{figure}[htp!]
\begin{center}
\setlength{\unitlength}{1cm}~\\[1em]
\begin{picture}(12,8)
\put(0,.7){\scalebox{1.2}[1.2]{\includegraphics*{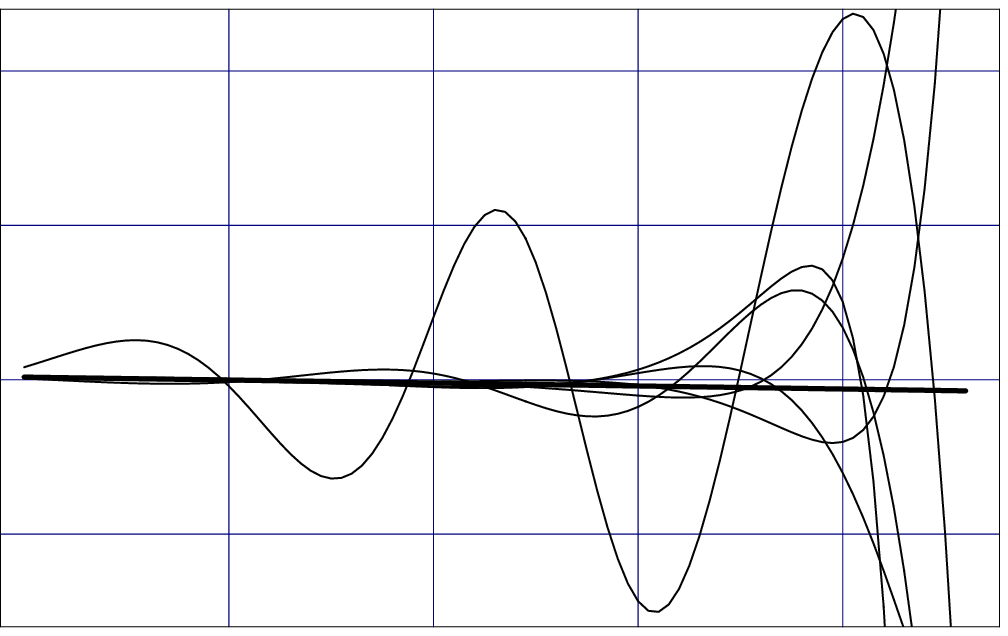}}}
\put(12.1,.3){$g$}
\put(2.2,.3){$-.014$}
\put(4.7,.3){$-.012$}
\put(7.2,.3){$-.01$}
\put(9.7,.3){$-.008$}
\put(-.7,1.8){$-.75$}
\put(-.6,3.67){$-.7$}
\put(-.7,5.54){$-.65$}
\put(-.6,7.4){$-.6$}
\put(10.9,7.9){A}
\put(11.45,7.9){B}
\put(10.26,1.9){C}
\put(10.35,2.6){D}
\put(11.1,.9){E}
\put(11.6,.9){F}
\end{picture}
\caption[NB-III]{Normalized imaginary part Im$[Z_{\rm var}^{(16)}(g)\exp{(-1/4g)}]$
 as a function of $g$ based on six different complex zeros (thin curves). The
fat curve represents the exact value, which is $Z_{\rm exact}(g)\simeq -0.7071+.524g-1.78g^2$. Oscillations of varying strength can be observed near $g=0$. Curves A
and C carry most smoothly near up to the origin.
Evaluation  based
on
 either of them yields equally good results.
 We have selected the zero belonging to curve C as our best choice to this order $L=16$.}
\label{NB-III}
\end{center}
\end{figure}

\begin{figure}[htp!]
\begin{center}
\setlength{\unitlength}{1cm}
\begin{picture}(19,5.5)
\put(0.5,0.5){\scalebox{.8}[.8]{\includegraphics*{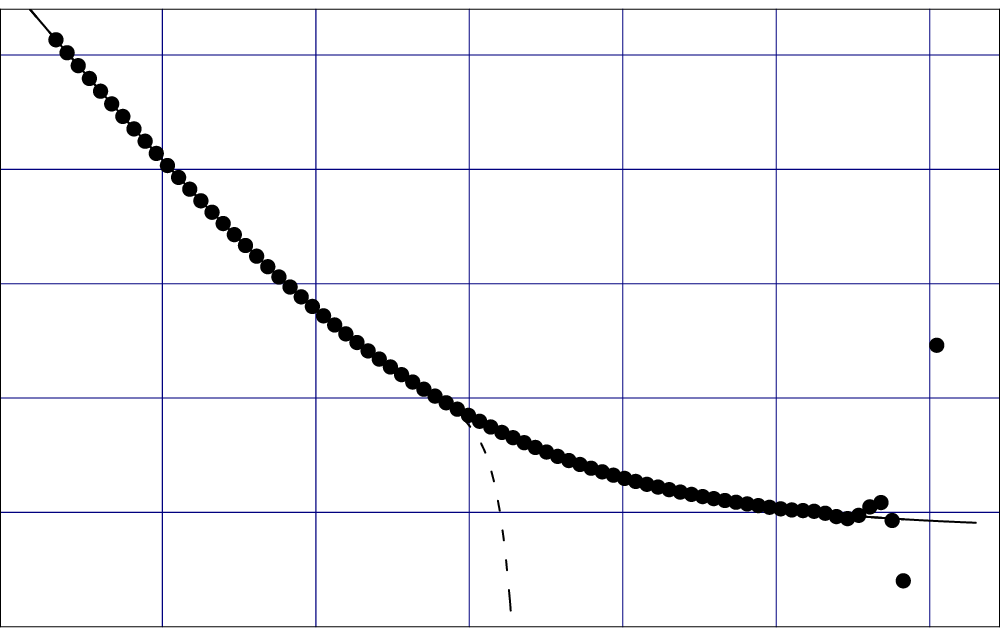}}}
\put(10,0.5){\scalebox{.8}[.8]{\includegraphics*{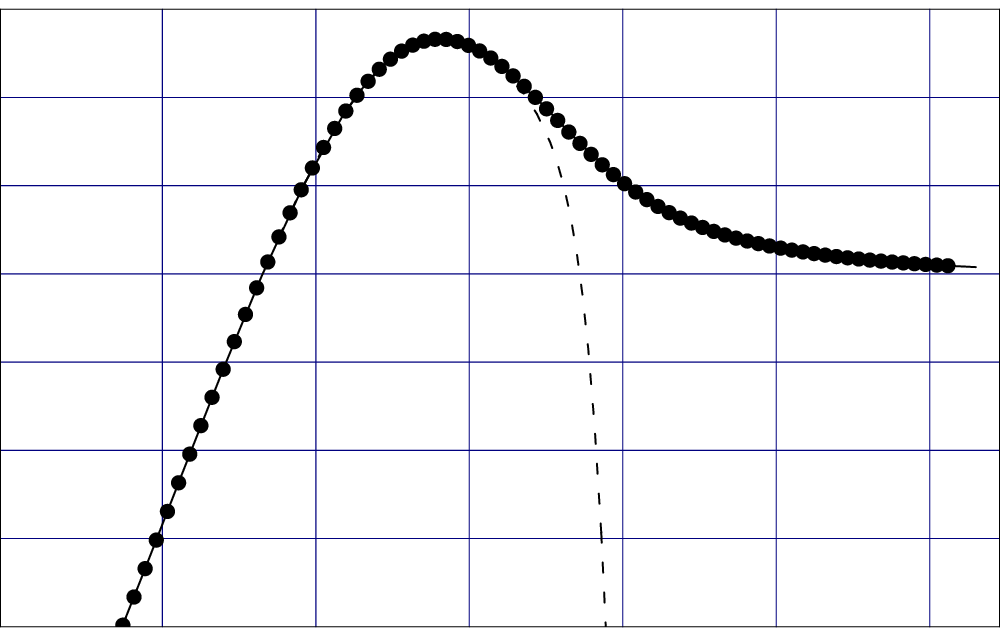}}}
\put(7.7,0.2){$\log{(-g)}$}
\put(1.72,0.2){$0$}
\put(4.,.2){$-2$}
\put(6.55,.2){$-4$}
\put(17.2,0.2){$\log{(-g)}$}
\put(11.22,0.2){$0$}
\put(13.55,.2){$-2$}
\put(16.05,.2){$-4$}
\put(9.55,1.9){$.9$}
\put(9.55,3.3){$1.0$}
\put(9.55,4.7){$1.1$}
\put(-.02,1.4){$-.7$}
\put(-.02,3.25){$-.6$}
\put(-.02,5.1){$-.5$}
\end{picture}
\caption[NB-IV]{Normalized imaginary part
Im$[Z_{\rm var}^{(16)}(g)\exp{(-1/4g)}]$ to the left
and the real part Re$[Z_{\rm var}^{(16)}(g)]$ to the right, based on the best zero C from Fig.~\ref{NB-III}, are plotted against $\log{|g|}$ as dots. The solid curve represents the exact function. The dashed curve is the 16th order of the strong-coupling expansion $Z^{(L)}_{\rm strong}(g)$ of equation (\ref{FP-STRONG}).}
\label{NB-IV}
\end{center}
\end{figure}

A better selection of the optimal $ \Omega $ values
comes from the following observation.
The
imaginary parts
of the approximations
near the singularity at $g=0$
show tiny
 oscillations.
The exact imaginary part is
known to decrease extremely fast,  like $\exp{(1/4g)}$, for $g \to 0-$,
practically without oscillations.
We can make  the tiny oscillations
more visible by taking
this exponential  factor out
of the imaginary part.
This is done in Fig.~\ref{NB-III}.
The oscillations differ
strongly  for
different choices of $ \Omega ^{(L)}$
 from the central region of the cluster. To each order $L$
we see that
one of them is smoothest in the sense that
the approximation
 approaches the singularity most closely before oscillations begin.
If this
$ \Omega ^{(L)}$
is chosen as the optimal one, we obtain excellent results
for the
entire
non-Borel-summable region $g<0$.
As an example, we pick the best zero for the $L=16$th order. Fig.~\ref{NB-III} shows the normalized imaginary part calculated to this order, but based on different zeros from the central cluster. Curve C appears optimal. Therefore we select the underlying zero as our best choice at order $L=16$
and calculate with it real
and imaginary part for the
non-Borel-summable region $-2<g<-.008$,
to be compared with the exact values.
Both are shown in Fig.~\ref{NB-IV}, where
we have again renormalized the imaginary part
by the exponential factor $\exp{(-1/4g)}$.
The agreement with the exact result (solid curve) is excellent
as was to be expected because of the fast convergence
observed in Fig.~\ref{NB-I}. It is
indeed much better than the strong-coupling expansion
to the same order, shown as a  dashed curve.
This is the essential improvement of our present theory as compared
to previously known methods probing
 into the tunneling regime \cite{Tunn}.

This non-Borel-summable regime will now be
investigated for the
quantum-mechanical
 anharmonic oscillator.
\section{
Tunneling Regime
of Quantum-Mechanical
 Anharmonic Oscillator}
\label{@sec3}
\begin{figure}[htp!]
\begin{center}
\setlength{\unitlength}{1.4cm}
\begin{picture}(10,6)
\put(.5,.4){\scalebox{1.12}[1.12]{\includegraphics*{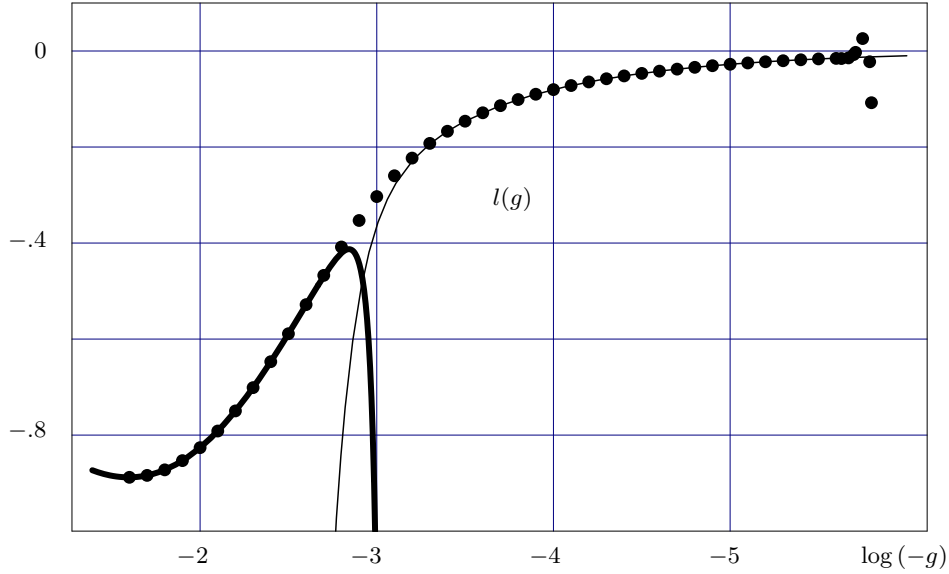}}}
\put(-.1,1.3){$\small {-.8}$}
\put(-.1,3.1){$\small {-.4}$}
\put(4.5,3.5){$l(g)$}
\put(.15,4.9){$\small {0}$}
\put(1.5,.1){$\small {-2}$}
\put(3.15,.1){$\small {-3}$}
\put(4.85,.1){$\small {-4}$}
\put(6.55,.1){$\small {-5}$}
\put(8.,.1){$\log{(-g)}$}
\end{picture}
\caption[I]{Logarithm of
 the imaginary part of
 the ground state energy of
 the anharmonic oscillator with the essential singularity
factored out for better visualization,
$
l(g)=\log\left[ {\sqrt{-\pi g/2}~E_{0,\rm var}^{(64)}(g)}\right] -1/3g$,
plotted against
 small negative values of
 the coupling constant $-0.2<g<-.006$
where the series is
 non-Borel-summable.
The thin curve represents
the divergent expansion around a
critical bubble
of Ref.~\cite{ZINNJ}.
The fat curve is the
 $22$nd order approximation of the
 strong-coupling expansion,
 analytically continued
to negative $g$ in the sliding regime calculated
 in
Chapter 17 of the textbook \cite{PI}.
 }
\label{I}
\end{center}
\end{figure}

The divergent weak-coupling perturbation expansion for the ground state energy of the anharmonic oscillator in the potential $V(x)=x^2/2+g\,x^4$ to order $L$
\begin{align}
\label{WEAK}
E_{0,\rm weak}^{(L)}(g)
 = \sum_{l=0}^L\; a_l\;g^l\,,
\end{align}
where $a_l=(1/2,\ 3/4,\ -21/8,\ 333/16,\ -30885/128,\ \dots)$, is
non-Borel-summable for $g<0$. It may be treated in the same way
as $Z(g)$ of the previous  model, making use as before
of Eqs.~(\ref{FP-VAR})--(\ref{FP-DERIV}), provided we set $p=1$
and $\omega=2/3$, so that $q=3$, accounting for the correct
power behavior $E_0(g)\propto g^{1/3}$ for $g \to \infty$. According to the principle of minimal dependence
and oscillations, we pick a best zero for the order $L=64$ from the
cluster of zeros of $P_L( \sigma )$,
and use it to calculate the logarithm of the normalized imaginary part:
\begin{align}
\label{IM64}
f(g):= \log{\left[ \sqrt{-\pi g/2}\ E_{0,\rm var}^{(64)}(g)\right] }-1/3g\,.
\end{align}
This quantity is plotted
in Fig.~\ref{I}
against $\log (-g)$
close to the tip of the left-hand cut
for $-.2<g<-.006$.
\noindent{}
Comparing our result to older values from
semi-classical calculations \cite{ZINNJ}
\begin{align}
\label{ZJ}
f(g)&=b_1 g-b_2 g^2+b_3 g^3-b_4 g^4+\dots ,
\end{align}
with
\begin{align}
\label{ZJ1}
 b_1 &=3.95833 \quad b_2=19.344 \quad b_3=174.21 \quad b_4=2177\ ,
\end{align}
 shown in Fig.~\ref{I} as
a thin curve, we find very good agreement.
This
expansion
contains the information
on the fluctuations
around the critical bubble.
It is divergent and
non-Borel-summable for $g<0$.
In Appendix A we have rederived
it in a novel way
which allowed us to extend and improve
it considerably.

Remarkably, our theory allows us
 to retrieve
the first three
terms of this expansion
from the
perturbation expansion.
Since our result
provides us with  a regular approximation
to
the essential singularity,
  the fitting procedure
 depends somewhat on
 the
 interval over which we fit
our curve by a power series.
A compromise between a sufficiently long
 interval and
the runaway of
 the divergent
 critical-bubble expansion is obtained for
a lower limit
 $g>-.0229\pm .0003$
and an
 upper limit $g=-0.006$. Fitting a polynomial to
 the data, we extract
 the following
 first three coefficients:
\begin{align}
\label{INST}
b_1 &=3.9586\pm .0003 \quad b_2=19.4\pm .12 \quad b_3=135\pm 18\ .
\end{align}
The agreement
of these numbers
 with those in
(\ref{ZJ})
demonstrates
 that
our method is capable of probing
 deeply into
 the
 critical-bubble region of the coupling constant.
\\

Further evidence
for the quality
of our theory
comes from a comparison
with  the analytically continued strong-coupling result
plotted to order $L=22$ as a
fat curve in Fig.~\ref{I}.
This expansion was derived by a
procedure
of summing non-Borel-summable series  developed
in Chapter 17
of the textbook \cite{PI}.
It was based on a two-step process:
the derivation of a strong-coupling
expansion of the type
(\ref{FP-STRONG}) from the divergent weak-coupling expansion,
and an
analytic
continuation of the strong-coupling expansion to negative $g$.
This method was applicable
only for large enough
coupling strength
where the
strong-coupling expansion converges,
the so-called {\em sliding regime\/}. It could
  not invade into the
 tunneling regime
at
small $g$
governed
by critical bubbles, which
 was treated
in \cite{PI}
 by a separate variational procedure.
The present
work fills the missing gap
by
extending variational perturbation theory
to {\em all\/} $g$ arbitrarily close to zero, without the need
 for a separate treatment of the tunneling regime.
\begin{figure}[htp!]
\begin{center}
\setlength{\unitlength}{1cm}
\begin{picture}(12,8)
\put(0,.7){\scalebox{1.2}[1.2]{\includegraphics*{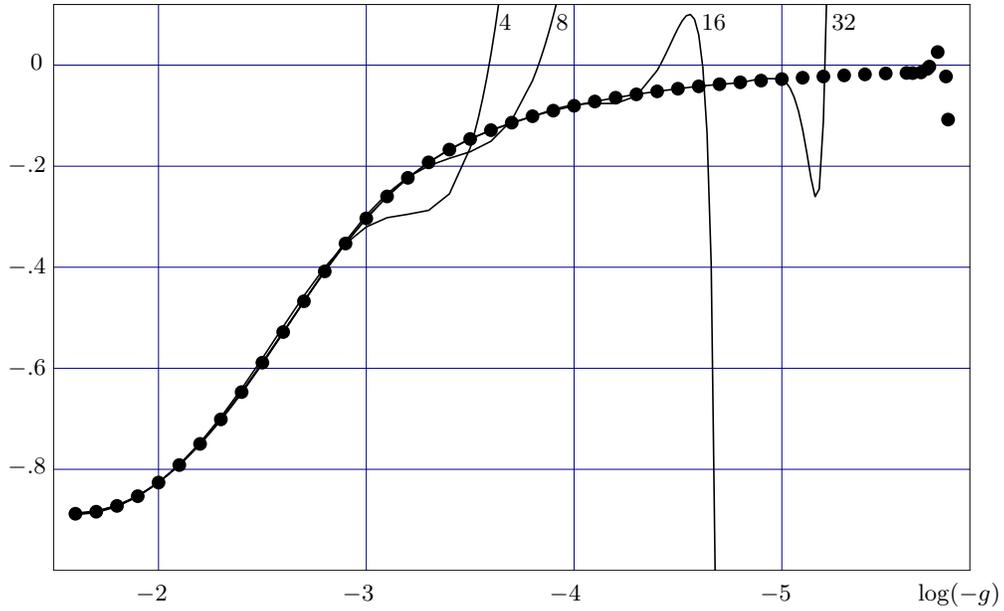}}}
\put(11.5,.3){$\log (-g) $}
\put(1.1,.3){$-2$}
\put(3.85,.3){$-3$}
\put(6.6,.3){$-4$}
\put(9.4,.3){$-5$}
\put(-.6,2){$-.8$}
\put(-.6,3.3){$-.6$}
\put(-.6,4.65){$-.4$}
\put(-.6,6){$-.2$}
\put(-.3,7.37){$0$}
\put(6.69,7.9){$8$}
\put(10.35,7.9){$32$}
\put(5.93,7.9){$4$}
\put(8.62,7.9){$16$}
\end{picture}
\caption[II]{Logarithm of the normalized imaginary part of the
ground state energy $\log{(\sqrt{-\pi g/2}\ E_{0,\rm var}^{L)}(g))}-1/3g$,
 plotted against $\log{(-g)}$ for orders $L=4,\ 8,\ 16,\ 32$ (curves). It is compared with the corresponding results for $L=64$ (points). This is shown for small negative values of the coupling constant $-0.2<g<-.006$, i.e. in the
non-Borel-summable critical-bubble region. Fast convergence is easily recognized. Lower orders oscillate more heavily. Increasing orders allow closer approach to the singularity at $g=0-$. }
\label{II}
\end{center}
\end{figure}

It is interesting to see, how the correct limit is approached as the order $L$ increases. This is shown in Fig.~\ref{II}, based on the
optimal zero in each order. For large negative $g$,
 even the small orders give excellent results.
Close to the singularity the scaling factor $\exp{(-1/3g)}$
will always win over
 the perturbation results.
It is surprising, however, how
fantastically close to the singularity we can go.

\section{Dynamic Approach to the Critical-Bubble Regime}
\label{@sec4}
Regarding the computational challenges connected with the
critical-bubble regime of small $g<0$, it is
 worth to develop an independent
 method to calculate imaginary parts
in the tunneling regime.
For a quantum-mechanical system
with an interaction potential
$g\,V(x)$,
such as a
the harmonic oscillator,
we may study the
effect of an  infinitesimal increase in $g$
upon the system.
It  induces an infinitesimal unitary transformation of the
 Hilbert space. The new Hilbert space can be made
 the starting point for the next infinitesimal increase in $g$.
In this way we derive an infinite set of first order ordinary
differential equations for the change of the energy levels
and matrix elements (for details see Appendix B):
\begin{figure}[htp!]
\begin{center}
\setlength{\unitlength}{1cm}
\begin{picture}(12,8.5)
\put(0,.7){\scalebox{1.2}[1.2]{\includegraphics*{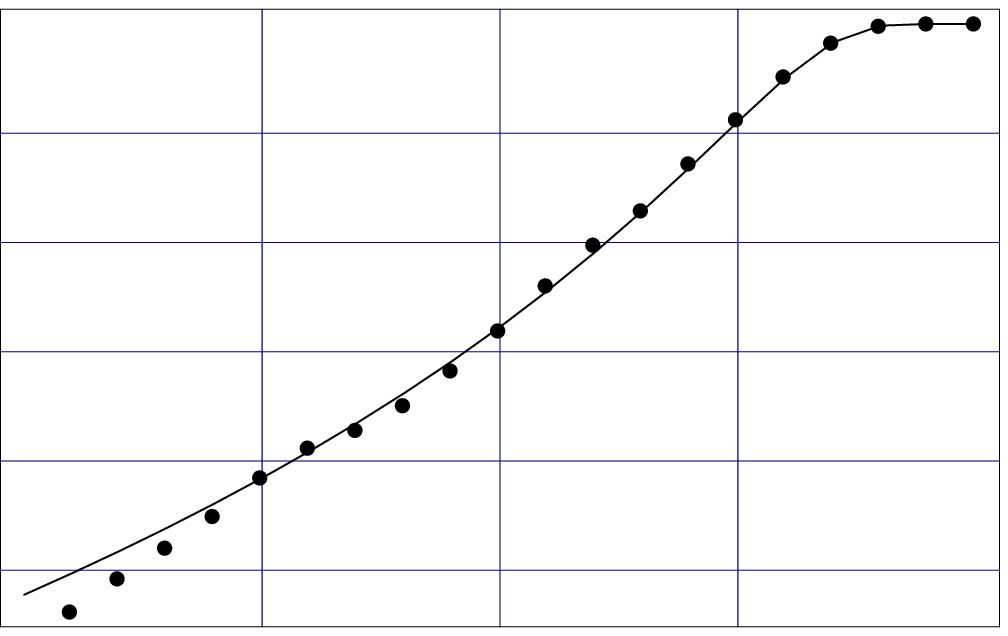}}}
\put(12,.3){$g$}
\put(2.85,.3){$-.3$}
\put(5.75,.3){$-.2$}
\put(8.6,.3){$-.1$}
\put(-.6,2.7){$-.2$}
\put(-.6,5.35){$-.1$}
\end{picture}
\caption[VIx]{Imaginary part of the ground state energy of
the anharmonic oscillator as solution of the coupled set of
differential equations (\ref{DYN}), truncated at the energy level of $n=64$ (points), compared with the corresponding quantity from the $L=64$th order of
non-Borel-summable  variational perturbation theory (curve), both shown as functions of the coupling constant $g$.}
\label{VIx}
\end{center}
\end{figure}
\begin{align}
\label{DYN}
E'_n(g) =& V_{nn}(g), \\
V'_{mn}(g) =& \sum_{k\neq n}\frac{V_{mk}(g)V_{kn}(g)}{E_m(g)-E_k(g)}+ \sum_{k\neq m}\frac{V_{mk}(g)V_{kn}(g)}{E_n(g)-E_k(g)}.
\end{align}
This system of equations
holds for  any one-dimensional Schroedinger problem.
Individual differences  come from  the
initial conditions,
which are the energy levels $E_n(0)$ of the unperturbed system
and the matrix elements $V_{nm}(0)$ of the interaction $V(x)$ in the unperturbed basis. For a numerical integration of the system a truncation is necessary. The obvious way is to restrict the Hilbert
space to the manifold spanned by the lowest $N$ eigenvectors of the unperturbed system. For cases like the anharmonic oscillator, which are even, with even perturbation
and with only an even state to be investigated, we may span the Hilbert space by even basis vectors only. Our initial conditions are thus for $n=0,\ 1,\ 2,\ \dots,\ N/2$:
\begin{align}
\label{INI}
E_{2n}(0) =& 2n+1/2 \\
V_{2n,2m} =& 0 \quad \mbox{if } m<0 \; \mbox{or } m>N/2 \\
V_{2n,2n}(0) =& 3(8n^2 + 4n + 1)/4\\
V_{2n,2n\pm 2}(0) =& (4n+3)\sqrt{(2n+1)(2n+2)}/2 \\
V_{2n,2n\pm 4}(0) =& \sqrt{(2n+1)(2n+2)(2n+3)(2n+4)}/4 \\
\end{align}
For the anharmonic oscillator with a $V(x)=x^4$ potential,
 all sums in equation (\ref{DYN}) are finite with at most four terms
due to the near-diagonal structure of the perturbation.

In order to find a solution for some $g<0$, we first integrate
the system from $0$ to $|g|$,
then around a semi-circle $g=|g|\exp{(i\varphi)}$
 from $\varphi=0$ to $\varphi=\pi$.
The imaginary part of $E_0(g)$ obtained
from a calculation with $N=64$ is shown in
Fig.~\ ref{VIx}, where it is
 compared with the variational result for $L=64$. The
agreement is excellent.
 It must be noted, however,
that the necessary truncation of the system of differential
equations introduces an error,
which cannot be made arbitrarily small by
increasing the truncation limit $N$. The approximations
are
asymptotic
sharing this property with the original weak-coupling series.
Its divergence is, however,
reduced considerably,
which  is the reason
why we obtain
accurate results
for the
critical-bubble regime, where
 the weak-coupling series
fails completely to
reproduce
the imaginary part.

\section{Appendix A}
We determine the ground state energy function $E_0(g)$ for the anharmonic oscillator on the cut, i.e. for $g<0$ in the bubble region, from the weak coupling coefficients $a_l$ of equation (\ref{WEAK}). The behavior of the $a_l$ for large $l$ can be cast into the form
\begin{align}
a_l/a_{l-1}=-\sum_{j=-1}^L \beta_j\ l^{-j}\ .
\label{B1}
\end{align}
The $\beta_j$ can be determined by a high precision fit to the data in the large $l$ region of $250<l<300$ to be
\begin{align}
\label{B2}
\beta_{-1,\ 0,\ 1,\ \dots}=&\left\{\ 3,\  -\frac{3}{2} ,\ \frac{95}{24},
  \ \frac{113}{6},\ \frac{391691}{3456},
  \ \frac{40783}{48},\ \frac{1915121357}{248832},
 \  \frac{10158832895}{124416},
 \  \frac{70884236139235}{71663616}, \right. \\& \nonumber
 \quad\  \frac{60128283463321}{4478976},
 \  \frac{286443690892}{1423},
 \  \frac{144343264152266}{43743},
 \  \frac{351954117229}{6},
 \  \frac{2627843837757582}{2339},\\& \nonumber \left.
 \quad  \frac{230619387597863}{10},
 \  \frac{12122186977970425}{24},
 \  \frac{41831507430222441029}{3550},\ \dots \right\}\ ,
\end{align}
where the rational numbers up to $j=6$ are found to be exact, whereas the higher ones are approximations.\\
Equation (\ref{B1}) can be read as recurrence relation for the coefficients $a_l$. Now we construct an ordinary differential equation for $E(g):=E_{\rm 0,weak}^{(L)}(g)$ from this recurrence relation and find:
\begin{align}
\left[\left(g \frac{d}{dg}\right)^L+g\sum_{j=0}^{L+1}\beta_{L-j} \left(g \frac{d}{dg}+1\right)^j \right]E(g)=0\ .
\label{B3}
\end{align}
All coefficients being real, real and imaginary part of $E(g)$ each have to satisfy this equation separately.
The point $g=0$, however,  is not a regular point. We are looking for a solution, which is finite when approaching it along the negative real axis. Asymptotically $E(g)$ has to satisfy $E(g)\simeq \exp{(1/g\beta_{-1})}= \exp{(1/3g)}$. Therefore we solve (\ref{B3}) with the ansatz
\begin{align}
E(g)=g^\alpha \ \exp{\left( \frac{1}{3g} -\sum_{k=1} b_k (-g)^k \right)}
\label{B4}
\end{align}
to obtain $\alpha = -1/2$ and
\begin{align}
\label{B5}
b_{1,2,3,\dots}=&\left\{\; \frac{95}{24},\ \frac{619}{32}\right.
     ,\ \frac{200689}{1152},
  \ \frac{2229541}{1024} ,
  \ \frac{104587909}{3072},
  \ \frac{7776055955}{12288}  ,
  \ \frac{9339313153349}{688128},
 \ \frac{172713593813181}{524288},\\ & \nonumber
     \quad\ \frac{1248602386820060039}
   {139886592},\ \frac{
       14531808399402704160316631}{
       54391637278720},
 \ \frac{12579836720279641736960567921}
   {1435939224158208},\\ & \nonumber
\left. \quad \frac{109051824717547897884794645746723}{348951880031797248},\ \frac{45574017678173074497482074500364087}{3780312033677803520}\ \dots
\right\}\ .
\end{align}
This is in agreement with equation (\ref{ZJ1}) and an improvement compared to the WKB results of \cite{ZINNJ}. Again, the first six rational numbers are exact, followed by approximate ones.
\section{Appendix B}
\noindent
Given a one-dimensional quantum system
\begin{align}
(H_0+g\ V)|n,g\rangle=E_n(g)|n,g\rangle
\label{A1}
\end{align}
with Hamiltonian $H=H_0+g\ V$, eigenvalues $E_n(g)$ and eigenstates $|n,g\rangle$ we consider an infinitesimal increase $dg$ in the coupling constant $g$. The eigenvectors will undergo a small change:
\begin{align}
|n,g+dg\rangle=|n,g\rangle+dg\ \sum_{k\ne n}u_{nk}|k,g\rangle
\label{A2}
\end{align}
so that
\begin{align}
\frac{d}{dg}|n,g\rangle=\sum_{k\ne n}u_{nk}|k,g\rangle\ .
\label{A3}
\end{align}
Given this, we take the derivative of (\ref{A1}) with respect to $g$ and multiply by $\langle m,g|$ from the left to obtain:
\begin{align}
\langle m,g|V-E'_n(g)|n,g\rangle=\sum_{k\ne n}u_{nk}\langle m,g|H_0+g\ V-E_n(g)|k,g\rangle\ .
\label{A4}
\end{align}
Setting now $m=n$ and $m \ne n$ in turn, we find:
\begin{align}
E'_n(g)=&V_{nn}(g)
\label{A5}\\
V_{mn}(g)=&u_{nm}\ \left(E_m(g)-E_n(g)\right)\ ,
\label{A6}
\end{align}
where $V_{mn}(g)=\langle m,g|V|n,g\rangle$.\\
Equation (\ref{A5}) governs the behavior of the eigenvalues as functions of the coupling constant $g$. In order to have a complete system of differential equations, we must also determine how the  $V_{mn}(g)$ change, when $g$ changes. With the help of equations (\ref{A3}) and (\ref{A6}), we obtain:
\begin{align}
V'_{mn}=&\sum_{k\ne m}u^*_{mk}\langle k,g|V|n,g\rangle + \sum_{k\ne n}u_{nk}\langle m,g|V|k,g\rangle
\label{A7} \\
V'_{mn}=& \sum_{k\ne m} \frac{V_{mk}V_{kn}}{E_m-E_k} +  \sum_{k\ne n} \frac{V_{mk}V_{kn}}{E_n-E_k} \ .
\label{A8}
\end{align}
Equations  (\ref{A5}) and (\ref{A8}) together describe a complete set of differential equations for the energy eigenvalues $E_n(g)$ and the matrix-elements $V_{nm}(g)$. The latter determine via (\ref{A6}) the expansion coefficients $u_{mn}(g)$. Initial conditions are given by the eigenvalues $E_n(0)$ and the matrix elements  $V_{nm}(0)$ of the unperturbed system.
\end{document}